\documentclass[]{emulateapj}

\begin{document}

\title{Limits on Very High Energy Emission from Gamma-Ray Bursts with the Milagro Observatory}

\author{R.~Atkins,\altaffilmark{1} W.~Benbow,\altaffilmark{2,9} D.~Berley,\altaffilmark{3} E.~Blaufuss,\altaffilmark{3} 
J.~Bussons,\altaffilmark{3,10} D.~G.~Coyne,\altaffilmark{2} 
T.~DeYoung,\altaffilmark{2,3} 
B.~L.~Dingus,\altaffilmark{7} D.~E.~Dorfan,\altaffilmark{2} R.~W.~Ellsworth,\altaffilmark{4} 
L.~Fleysher,\altaffilmark{6} R.~Fleysher,\altaffilmark{6} G.~Gisler,\altaffilmark{7} M.~M.~Gonzalez,\altaffilmark{1} 
J.~A.~Goodman,\altaffilmark{3} T.~J.~Haines,\altaffilmark{7} E.~Hays,\altaffilmark{3} C.~M.~Hoffman,\altaffilmark{7} 
L.~A.~Kelley,\altaffilmark{2} 
J.~E.~McEnery,\altaffilmark{1,11} R.~S.~Miller,\altaffilmark{5}
A.~I.~Mincer,\altaffilmark{6} M.~F.~Morales,\altaffilmark{2,12,14} P.~Nemethy,\altaffilmark{6} D.~Noyes,\altaffilmark{3} 
J.~M.~Ryan,\altaffilmark{5} F.~W.~Samuelson,\altaffilmark{7} 
A.~Shoup,\altaffilmark{8} G.~Sinnis,\altaffilmark{7} A.~J.~Smith,\altaffilmark{3} G.~W.~Sullivan,\altaffilmark{3} 
D.~A.~Williams,\altaffilmark{2} S.~Westerhoff,\altaffilmark{2,13} 
M.~E.~Wilson,\altaffilmark{1} X.~Xu\altaffilmark{7} and G.~B.~Yodh\altaffilmark{8}}

\altaffiltext{1}{University of Wisconsin, Madison, WI 53706}
\altaffiltext{2}{University of California, Santa Cruz, CA 95064}
\altaffiltext{3}{University of Maryland, College Park, MD 20742}
\altaffiltext{4}{George Mason University, Fairfax, VA 22030}
\altaffiltext{5}{University of New Hampshire, Durham, NH 03824-3525}
\altaffiltext{6}{New York University, New York, NY 10003}
\altaffiltext{7}{Los Alamos National Laboratory, Los Alamos, NM 87545}
\altaffiltext{8}{University of California, Irvine, CA 92717}
\altaffiltext{9}{Now at Max Planck Institute, Heidelberg, Germany}
\altaffiltext{10}{Now at Universite de Montpellier II, Montpellier, France}
\altaffiltext{11}{Now at NASA Goddard Space Flight Center, Greenbelt, MD 20771}
\altaffiltext{12}{Now at Massachusetts Institute of Technology, Cambridge, MA 02139}
\altaffiltext{13}{Now at Columbia University, New York, NY 10027}
\altaffiltext{14}{Corresponding author: M. F. Morales, (617) 452-3159, mmorales@space.mit.edu}

\begin{abstract}
The Milagro telescope monitors the northern sky for 100 GeV to 100 TeV transient emission through continuous very high energy wide-field observations.  The large effective area and $\sim$100 GeV energy threshold of Milagro allow it to detect very high energy (VHE) gamma-ray burst emission with much higher sensitivity than previous instruments, and a fluence sensitivity at VHE energies comparable to that of dedicated gamma-ray burst satellites at keV to MeV energies.  Even in the absence of a positive detection, VHE observations can place important constraints on  gamma-ray burst (GRB) progenitor and emission models. We present limits on the VHE flux of 40 s -- 3 h duration transients nearby to earth, as well as sensitivity distributions which have been corrected for gamma-ray absorption by extragalactic background light and cosmological effects.  The sensitivity distributions suggest that the typical intrinsic VHE fluence of GRBs is similar or weaker than the keV -- MeV emission, and we demonstrate how these sensitivity distributions may be used to place observational constraints on the absolute VHE luminosity of gamma-ray bursts for any GRB emission and progenitor model.
\end{abstract}


\section{Introduction}

Very high energy gamma-ray burst (GRB) observations have the potential to constrain theoretical models of both the prompt and extended phases of GRB emission. Models based on both internal and external shocks have predicted VHE fluence comparable to, or in certain situations stronger than, the keV -- MeV radiation, with durations ranging from shorter than the keV -- MeV burst to extended TeV afterglows \citep{Dermer:VHEGRB,PillaLoeb:VHEGRB,Bing}.  GRB emission above 100 GeV is particularly sensitive to the Lorentz factor and the photon density of the emitting material --- and thus the distance of the radiating shock from the source --- due to $\gamma\gamma \rightarrow e^+e^-$ absorption in the emission region.  

EGRET observed several GRBs at GeV energies and observed no evidence of a high energy rollover in the GRB spectrum \citep{EGRET}, and recent results by \citet{Gonzalez} indicate that the spectrum of some GRBs contains a very hard, luminous, long duration component.  At energies above a few 100 GeV several observation attempts have been made, including: coincidence observations between the Tibet air-shower array and BATSE \citep{TibetGRB}, rapid follow-up observations by the Whipple Air-Cherenkov telescope \citep{WhippleGRB}, and coincident and monitoring studies by AIROBICC \citep{Padilla}, Whipple \citep{WhippleGRB1sec}, and the Milagro prototype \citep{MilagritoGRB, Leonor, McCullough}.  While several of these studies have reported evidence for VHE GRB emission, no clear picture of the VHE emission from gamma-ray bursts has emerged.


In recent years the rate of possible GRB coincident detections has been relatively low due to the absence of sensitive, large field of view GRB detectors like BATSE. However, Milagro is a very sensitive detector with a low $\sim$100 GeV energy threshold and wide $\gtrsim$ 2 sr field of view, and is fully capable of autonomous identification of a VHE burst of emission. The low energy threshold is particularly important because of the reduced attenuation by extragalactic background light near 100 GeV \citep{Jelley,Primack:TeVAbsorption,Stecker:EBL}, which dramatically increases the volume of space observed. This paper details the monitoring of the northern sky for VHE transients of 40 s -- 3 hours duration with the Milagro observatory.

\section{The 40 s to 3 hour Transient Search in Milagro}
\label{40sSearch}

Very high energy gamma rays incident on the earth interact in the upper atmosphere and produce extensive air showers (EAS) which propagate to lower altitudes.  Milagro uses the water Cherenkov technique to detect EASs by converting the wave front of relativistic particles in the EAS into a front of Cherenkov light, detected by photomultiplier tubes (PMTs) \citep{GritoNIM}. The amplitude and arrival time of the PMT signals are analyzed to determine the direction and characteristics of the $\sim$1,800 EAS per second detected by Milagro, most of which are initiated by cosmic ray nuclei. The water Cherenkov technique enables Milagro to operate continuously. 

The precision of the Milagro event reconstruction depends strongly on the shower characteristics, leading to a point spread function (PSF) for the initiating particle direction which varies considerably from event to event.  The arrival times measured by the PMTs are fit to a plane, after making corrections based on the pulse amplitude in each PMT and the distance of the PMT from the reconstructed shower core.  The shower fit proceeds through several iterations, at each stage removing times with large residuals to the fit shower plane. Showers reconstructed within $45^{\circ}$ of zenith are kept for further analysis.  The number of PMT times surviving in the final iteration is called $n_{\rm fit}$, which is used in conjunction with the reduced $\chi^{2}$ to characterize the quality of the reconstruction. The showers are then separated into 13 classes, based on the values of $n_{\rm fit}$ and $\chi^2$, such that events within each class have similar angular PSFs (see \citet{MyThesis} for details).  In general, the width of the PSF decreases as $n_{\rm fit}$ increases or $\chi^2$ decreases.  The class with the worst PSF is removed from further analysis, at a negligible cost in sensitivity, because the PSF was so broad that it could not be readily handled by the code used to accumulate the sky maps.

The characteristics of the light recorded by PMTs at the bottom of the reservoir can be used to reject background cosmic ray showers in favor of gamma
rays using a quantity we call ``compactness'' \citep{CrabPaper}.  Events are assigned a weight, between 0 and 1, based on the relative probability of the event being a background cosmic ray or a gamma ray, given the compactness value and to which of the 12 remaining PSF classes the event belongs. To speed the computation, events with weight less than 0.5 are rejected.

This search implements an analysis method developed by \citet{WATPaper} which efficiently includes the event-by-event PSF and photon probability (i.e. background rejection) information, enhancing the sensitivity of the observations.  The technique uses
the characteristics described above to determine a spatial photon probability density --- the expected PSF of the event multiplied by the probability the shower was initiated by a gamma ray --- which is then added to a sky map to form the estimated total photon density.  Therefore, rather than considering each photon as a point at each reconstructed event location, the new technique instead places a smooth distribution at that location, with the shape of the distribution mirroring the reconstruction accuracy of the detector.  
The probability of the background, which primarily consists of cosmic ray events, producing an observed peak in the sky map is then given by the expected statistical distribution of the total photon density at that location on the sky map.  The expected statistical distributions, which are typically not Gaussian in shape, can be calculated directly from the PSF characteristics of the detector and placed into a lookup table to enable very fast searches for unknown gamma-ray sources.  This technique is similar to maximum likelihood analyses in sensitivity, but is computationally fast enough to perform a real time transient search in a high data-rate detector like Milagro . 

The 40 s -- 3 h transient search is performed by first dividing the data into 20 second intervals and creating sky maps, both for the actual data (``on source") and for the expected background in each time interval.  From these initial maps, 40 s sky maps are formed by summing all pairs of successive 20 s maps.  The resulting set of sky maps is temporally oversampled by a factor of two, with every 20 s map used twice so that a 40 s burst is guaranteed to be largely contained in one of the overlapping 40 s maps.  A search is performed on each of these maps for emission using the test statistic described below, then pairs of the 40 s maps are combined to make 80 s maps ($1/2$ of the 40 s maps are discarded to maintain the factor of two oversampling), and so on. The nine specific time scales that were searched are 40, 80, 160, 320, 640, 1280, 2560, 5120, and 10240 seconds.  The total observation days for each time scale is shown in the second column of Table \ref{probtable}. The longer time scales have slightly less observation time due to gaps from occasional data dropouts.  Full details of the transient search can be found in \citet{MyThesis}.

The significance of excesses in the on-source sky maps, compared to the background sky maps, is evaluated using a test statistic. The test statistic is very closely related to the probability of an observation being produced by a background fluctuation, with approximations used to speed up the analysis breaking the exact one-to-one relationship and leading to a slight flattening of the test statistic distributions \citep{WATPaper}.  Since the value of the test statistic decreases with increasing transient flux (lower probability of the observation being produced by a background fluctuation), a discovery search simply looks for values of the test statistic below that expected from the number of independent observations.  If a transient is identified, the statistical significance can be calculated from the test statistic.

\begin{figure}
\plotone{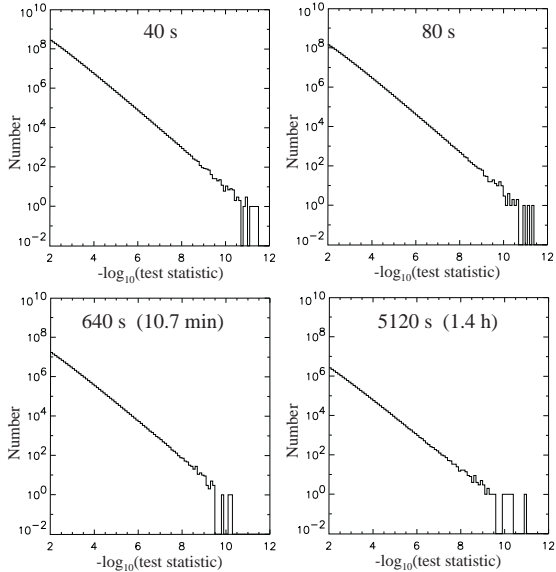}
\caption{Test statistic distributions observed by Milagro between 2001 May $2^{\rm nd}$ and 2002 May $22^{\rm nd}$ for 4 representative time scales. No evidence for transient VHE emission was observed. No test statistic below $10^{-12}$ was observed in any of the nine time scales studied.}
\label{totalResult}
\end{figure}

The data used for this analysis were taken between 2001 May $2^{\rm nd}$ and 2002 May $22^{\rm nd}$. Example distributions of the test statistic for 4 time scales are shown in Figure \ref{totalResult}, with no observed location for any of the nine search intervals having a test statistic below the $\sim$$10^{-12}$ value expected from the number of independent observations. There is no evidence for transient VHE emission of 40 s to 3 hours duration in the Milagro data between 2001 May $2^{\rm nd}$ and 2002 May $22^{\rm nd}$.

\section{Limits on VHE GRB Emission}
\label{LimitSection}

Two sets of constraints are presented here --- upper limits on the VHE flux at the earth, and sensitivity distributions which constrain the absolute VHE luminosity of gamma-ray bursts.  The flux limits have the advantage of being independent of cosmological models and directly comparable to previous VHE GRB limits.  However, due to the energy-dependent absorption of VHE photons by extragalactic background light (EBL), it is difficult to compare flux limits at the earth to theoretical calculations.  The sensitivity distributions place the observations into a theoretical context by correcting for absorption and cosmological effects.  In conjunction with model predictions these can be used to determine the maximum VHE luminosity at the source which is consistent with the observations presented in this paper.

Since no observation in any of the nine time scales had a test statistic below $10^{-12}$, we can use this as a threshold to set a conservative limit on the maximum flux consistent with these observations. The relationship between the gamma ray flux and the test statistic was determined using a combination of Monte Carlo generated transient VHE signals and real data background.  The simulated detector signal was modeled from 100 GeV to 21 TeV as a function of zenith angle and observed spectrum, propagated through the Milagro reconstruction code, added to representative data, and fed to the 40 s to 3 hour transient detection program. If the test statistic for the simulated signal fell below the $10^{-12}$ threshold of this study, the signal would have been stronger than any observed fluctuation on any time scale. 

\begin{figure}
\plotone{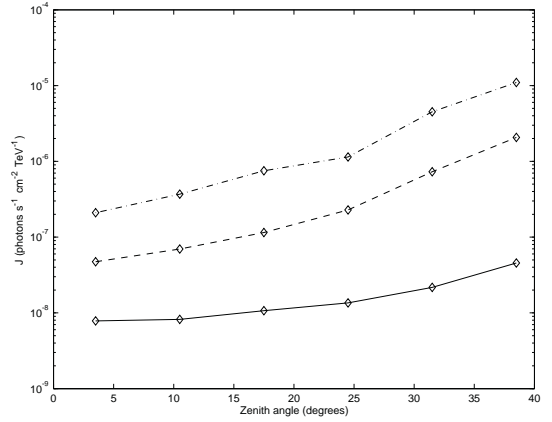}
\caption{The 90\% confidence upper limits for VHE transients of 40 s duration and an $E^{-2.0}$ spectrum for three different values of a hard spectral cutoff. The diamonds indicate the calculated limits on the normalization factor $J$ in photons s$^{-1}$ cm$^{-2}$ TeV$^{-1}$ for the spectrum $\frac{dN}{dE}=J(\frac{E}{1 {\rm TeV}})^{-2.0}$.  Lines connecting the points have been drawn to guide the eye, with a solid line for a 21 TeV cutoff, dashed line for 1 TeV, and dash-dot line for 300 GeV.  Sensitivity values for the eight longer time scales (80 s, 160 s, 320 s, 640 s, 1280 s, 2560 s, 5120 s, and 10240 s) are not plotted, but the sensitivity improves as a $J \propto t^{-0.64}$ power law with the duration of the VHE transient.  Monte Carlo statistics lead to an uncertainty in the upper limits of 19\%.  Systematic errors are due principally to uncertainties in the Monte Carlo simulation and are estimated to be +40\%/-20\%, for a total estimated error of +44\%/-27\%.}
\label{LocalLimit2.0}
\end{figure}

The flux limits represent the VHE transient signals which would have produced a signal with a test statistic below $10^{-12}$ $90\%$ of the time (signal from a single burst which is excluded at the $90\%$ confidence level). For this paper $E^{-2.0}$ power-law spectra extending from 100 GeV to hard spectral cutoffs at 300 GeV, 1 TeV, or 21 TeV (as observed local to the earth) were chosen.  These spectra serve as approximate models for an inverse Compton spectral bump at TeV energies or a hard VHE extension of the observed GRB spectrum past the multi-GeV observations of EGRET \citep{EGRET}, with the spectral cutoff approximating either an intrinsic cutoff for nearby sources or EBL absorption for more distant sources.
The limits are presented in Figure \ref{LocalLimit2.0} as flux density as a function of zenith angle and burst duration.\footnote{The full set of limits for $E^{-2.0}$ and $E^{-2.4}$ spectra without cutoffs are shown in \citet{MyThesis}.}  Given the $\sim0.8$ years of observation time
and the $\sim2$ sr field of view, these flux limits provide the strongest constraints to
date on the the number of nearby bursts which emit TeV photons.  For example, at a redshift of 0.024 (the 100 Mpc distance of the nearby bursts identified by \citet{NorrisNearBursts}, with the cosmology used in that paper) our 90\% confidence limit for a spectrum which extends to 21 TeV and burst emission which lasts 40 s at a 10.5 degree zenith angle is $1.1 \times 10^{46}$ ergs/s.

However, at larger distances the usefulness of the limits from Figure \ref{LocalLimit2.0} is complicated by the absorption of VHE photons by extragalactic background light \citep{Jelley,Primack:TeVAbsorption,Stecker:EBL}.  The observed spectrum is a convolution of the emitted spectrum and the extragalactic background light absorption
--- neither of which are well determined.  Limits on the absolute VHE luminosity of GRBs are thus necessarily model dependent.  In setting the limits which follow, Milagro is more sensitive than might initially be expected because bursts above $\sim$10$^{51}$ erg/s are sufficiently bright to be detected even after extinction of a few optical depths.

In an effort to place the current observations in context, a set of assumptions about the emitted spectrum, EBL absorption, and cosmology have been chosen and sensitivity distributions calculated within this theoretical framework.  For a set of source luminosities and durations, simulated GRBs were created with an $E^{-2.0}$ emission spectrum from 100 GeV to 21 TeV, an isotropic sky position, and following the star formation rate in z.  The EBL absorption was determined by \citet{BullockEBL} (similar to \citet{Primack:TeVAbsorption}) using recent results from semi-analytic modeling, and includes collisional starburst effects and a Kennicut initial mass function \citep{Kennicut}.\footnote{The magnitude of the EBL absorption is quite uncertain, and the models by \citet{Stecker:EBL} have considerably less attenuation.  Consequently the limits presented here may be conservative.} A $\Lambda$CDM cosmology ($\Omega_{\rm M} = .3$, $\Omega_{\Lambda} = .7$, ${\rm h}=0.65$) was used for both the EBL calculation and the cosmological corrections.  The results are rather insensitive to the emitted spectrum and cutoff energy due to the strong EBL absorption which eliminates nearly all emission above 300 GeV at redshift $\approx 0.3$, and limits the distance to which Milagro can observe VHE emission to redshift $\lesssim 0.7$.  The star formation rate (SFR) was modeled as $10^{1.45z - 1.52}$ M$_\odot$/Mpc$^3$/year  for $z<0.9$ and 0.61 M$_\odot$/Mpc$^3$/year for $z\geq0.9$ within each redshift bin, following the discussion in \citet{Hopkins}.

The results of the sensitivity calculation are presented in Table \ref{probtable} as the probability that Milagro would have observed a VHE GRB (i.e.\ test statistic $<10^{-12}$) as a function of the absolute luminosity,\footnote{The luminosity quoted is the isotropic luminosity assuming no beaming at the source, and is used to facilitate comparison with MeV to GeV measurements.} distance, and duration of the source.  Due to increased atmospheric depth away from detector zenith, the low energy threshold of Milagro increases with zenith angle, as can be seen in Figure \ref{LocalLimit2.0} by the decreased sensitivity at large zenith angles when the emission is limited to lower energies.  As GRBs become increasingly distant the EBL absorption effectively truncates the observed spectrum above a few hundred GeV.  These effects lead to the effective field of view of Milagro becoming narrower with increased redshift, producing the gradual decline in detection probability seen in Table \ref{probtable}.

Given theoretical expectations for the GRB distance, luminosity, and duration distributions, the probabilities in Table \ref{probtable} can be used to calculate the upper limits provided by these observations.  For a 90\% confidence upper limit, the maximum GRB rate consistent with these observations is given by multiplying the expected number of GRBs within each redshift bin by the detection probability in Table \ref{probtable}, and adjusting the total number of GRBs until the mean number of events seen by Milagro is 2.3.  If the model GRB distance distribution does not follow the SFR, the probabilities in Table \ref{probtable} can still be used as approximate values, or corrections can be applied to adjust for the distribution of events within each bin. The results are typically quoted as the number of GRBs/${\rm M}_\odot$ of star formation or the average number of GRBs/${\rm Gpc}^{3}\!/{\rm year}$.   

As an example, consider a model which predicts that GRBs follow the star formation rate, with all GRBs emitting a characteristic 80 s pulse of VHE emission.  The resulting 90\% confidence upper limits for this model are $6.2\times 10^{-8}$ GRBs/${\rm M}_\odot$ of star formation (an average of $4.8$ GRBs/${\rm Gpc}^{3}\!/{\rm year}$ over 0$<$z$<$0.5) for an isotropic luminosity of $10^{51}$ ergs/s, or $1.1\times 10^{-8}$ GRBs/${\rm M}_\odot$ of star formation (an average of 0.8 GRBs/${\rm Gpc}^{3}\!/{\rm year}$ over 0$<$z$<$0.5) for a luminosity of $10^{52}$ ergs/s, with significantly tighter constraints if the GRB distribution trails the star formation rate (i.e.\ there are more low redshift GRBs).  Of the thirty-six GRBs with known distances, five have a redshift below 0.5.\footnote{One can use this fraction to roughly estimate that if all bursts were 10$^{51}$ erg/s, they would yield a flux brighter than the sensitivity in  Figure \ref{LocalLimit2.0} ($\sim 10^{-7}$ cm$^{-2}$ s$^{-1}$) out to redshift of about 0.5 for zenith angles less than $\sim30^{\circ}$, and approximately $(5/36) \times 0.84/(4\pi) = 1\%$ of all bursts would be detectible by Milagro.}  If the GRBs detected by BATSE follow the same distance distribution, a rough estimate of the observed GRB rate yields $\sim$$2.6$ GRBs/${\rm Gpc}^{3}\!/{\rm year}$, or $\sim$$3.4\times 10^{-8}$ GRBs/${\rm M}_\odot$ of star formation if they follow the SFR.  While detailed model calculations are needed to convert the probabilities in Table \ref{probtable} into meaningful upper limits, comparison with the limits from this simple model suggests that if GRBs follow the star formation rate, the typical luminosity of 40 s -- 3 h VHE GRB counterparts is constrained to be similar to or less than the prompt keV -- MeV emission.

\section*{Acknowledgments}

We acknowledge Scott Delay and Michael Schneider for their dedicated efforts in the 
construction and maintenance of the Milagro experiment.  This work has been supported by the 
National Science Foundation (under grants 
PHY-0070927, 
-0070933, 
-0075326, 
-0096256, 
-0097315, 
-0206656, 
-0302000, 
and
ATM-0002744) 
the US Department of Energy (Office of High-Energy Physics and 
Office of Nuclear Physics), Los Alamos National Laboratory, the University of
California, and the Institute of Geophysics and Planetary Physics.  MFM was a
NASA Graduate Student Researcher.

\begin{deluxetable}{c|c|ccc|ccc|ccc}
\tabletypesize{\scriptsize}
\tablecaption{Probability of GRB detection by Milagro\label{probtable}. This table indicates the probability of a GRB of given duration, absolute luminosity, and redshift within the 1.84 sr field of view of this analysis being detected by the Milagro experiment.  L$_x$ indicates the absolute luminosity (not corrected for beaming) in $10^x$ ergs/s between 100 GeV and 21 TeV.   Since the star formation rate and space volume both increase with redshift, the median redshift of the events used to calculate each probability is not in the center their redshift bin, but instead much nearer the upper redshift value.  The triple dot symbols represent values which were not calculated due to computational constraints.}

\tablehead{& & \phn & L$_{50}$ & \phn & \phn & L$_{51}$ & \phn & \phn & L$_{52}$ & \phn}

\startdata
Search & Total Obs. & & z & & & z & & & z & \\
Interval & Duration &0--0.1&0.1--0.2&0.2--0.3&0--0.15&0.15--0.3&0.3--0.45&0--0.23&0.23--0.47&0.47--0.7\\
\tableline
40 s & 290.2 d & 0.90 & 0.37 & 0.01 & 1.00 & 0.44 & 0.03 & 1.00 & 0.37 & 0.003 \\
80 s & 290.2 d & 1.00 & 0.27 & 0.01 & 1.00 & 0.36 & 0.09 & 1.00 & 0.54 & 0.08\\
160 s & 290.0 d & 1.00 & 0.38 & 0.04 & 1.00 & 0.59 & 0.11 & 1.00 & 0.58 & 0.10\\
\tableline
 & && z & & & z & & & z & \\
 &&0--0.15&0.15--0.3&0.3--0.45&0--0.23&0.23--0.47&0.47--0.7& 0--0.23&0.23--0.47&0.47--0.7\\
\tableline
320 s & 289.7 d & 0.73 & 0.10 & 0.00 & 1.00 & 0.18 & 0.01 & 1.00 & 0.68 & 0.17\\
640 s & 289.2 d & 0.91 & 0.20 & 0.01 & 1.00 & 0.27 & 0.01 & 1.00 & 0.81 & 0.24\\
1,280 s & 288.1 d &1.00 & 0.28 & 0.02 & 1.00 & 0.31 & 0.03 & 1.00 & \nodata & \nodata\\
2,560 s & 286.1 d & 1.00 & 0.45 & 0.03 & 1.00 & 0.40 & 0.07 & 1.00 & \nodata & \nodata\\
5,120 s & 282.2 d & 1.00 & 0.47 & 0.05 & 1.00 & 0.48 & 0.08 & 1.00 & \nodata & \nodata\\
10,240 s & 275.7 d &1.00 & 0.51 & 0.13 & 1.00 & 0.54 & 0.13 & 1.00 & \nodata & \nodata\\
\enddata
\end{deluxetable}

\end{document}